\documentclass[UKenglish, twocolumn, amsmath, amssymb, prl, aps, floatfix, superscriptaddress, 10pt]{revtex4-2}

\DeclareUnicodeCharacter{0308}{}

\usepackage{listings}
\usepackage{color}
\usepackage{xcolor}
\usepackage{graphicx}
\usepackage{colortbl}
\usepackage{booktabs}
\usepackage{caption}

\usepackage{physics}
\usepackage{braket}
\usepackage{ulem}
\usepackage[T1]{fontenc}
\usepackage{mathrsfs}  
\usepackage[dvipsnames]{xcolor}
\usepackage{tabularray}
\usepackage{hyperref}

\definecolor{tocblue}{RGB}{0,0,120}
\definecolor{figred}{RGB}{200,0,0}

\hypersetup{
	colorlinks=true,
	linkcolor=figred,
	citecolor=green,
	urlcolor=figred
}

\usepackage{orcidlink} % ORCID icon package (RevTeX supports it)
\usepackage{soul}
\usepackage[uncertainty-mode = compact, separate-uncertainty-units = single,  bracket-unit-denominator = false]{siunitx}

\makeatletter

\newcommand{\listsubsectioninsi}{%
	\@starttoc{sitoc}%
}

\makeatother

\AtBeginDocument{\RenewCommandCopy\qty\SI} % avoid conflict with Physics
\usepackage{booktabs} % prettier tables
\usepackage{bm}

\newcommand{\cnn}{Universit\'e Paris-Saclay, Centre de Nanosciences et de Nanotechnologies, CNRS, 10 Boulevard Thomas Gobert, 91120, Palaiseau, France}
\newcommand{\quandela}{Quandela, 7 Rue Léonard de Vinci, 91300 Massy, France}
\newcommand{\nancy}{
	Université de Lorraine, CNRS, LPCT, F-54000 Nancy, France}
\newcommand{\majulab}{MajuLab, CNRS-UCA-SU-NUS-NTU International Joint Research Laboratory, 117543 Singapore, Singapore}
\newcommand{\cqt}{Centre for Quantum Technologies, National University of Singapore, 117543 Singapore, Singapore.}

\begin{document}
\title{Generation of quantum light with tunable Wigner negativity with a semiconductor quantum dot}

\author{Hubert Lam\,\orcidlink{0000-0002-1712-9201}}
\thanks{These authors contributed equally to this work.}
\affiliation{\cnn}
%\affiliation{These authors contributed equally to this work.}

\author{Petr Steindl\,\orcidlink{0000-0001-9059-9202}}
\thanks{These authors contributed equally to this work.}
\email{petr.steindl@universite-paris-saclay.fr}
\affiliation{\cnn}
%\affiliation{These authors contributed equally to this work.}

\author{Yann Portella\,\orcidlink{0009-0006-6362-9246}}
\affiliation{\cnn}

\author{Kiarn T. Laverick\,\orcidlink{0000-0002-3688-1159}}
\affiliation{\majulab}
\affiliation{\cqt}

\author{Kian Hwee Lim\,\orcidlink{0000-0003-2154-4288}}
\affiliation{\majulab} \affiliation{\cqt}

\author{Juan R. Álvarez\,\orcidlink{0000-0003-3411-8237}}
\affiliation{\cnn}

\author{Aristide Lema\^itre\,\orcidlink{0000-0003-1892-9726}}
\affiliation{\cnn}

\author{Maria Maffei\,\orcidlink{0000-0001-5183-4716}}
\affiliation{\nancy}
\author{Alexia Auffèves\,\orcidlink{0000-0003-4682-5684}}
\affiliation{\majulab}
\affiliation{\cqt}

\author{Dario A. Fioretto\,\orcidlink{0000-0003-3829-4000}}
\affiliation{\cnn} \affiliation{\quandela}
\author{Pascale Senellart\,\orcidlink{0000-0002-8727-1086}}
\email{pascale.senellart-mardon@universite-paris-saclay.fr}
\affiliation{\cnn}
	
%\keywords{Keyword1, Keyword2, Keyword3}
	
\begin{abstract}
    Wigner-negative states of light are key resources for quantum advantage in continuous-variable optical quantum computing. They can be produced probabilistically via frequency conversion and photon detection, or deterministically by exploiting the single-photon nonlinearity of a quantum emitter. Here, we generate such Wigner-negative states of light with a tunable negativity using a single semiconductor quantum dot in a cavity. Optical coherent control of the quantum-dot transition allows to generate a quantum superposition of zero and one photon. The generated quantum states are characterized point-by-point in the phase space through homodyne measurement. The emergence of Wigner-function negativity is observed as the excitation power is increased. After correcting for collection and propagation losses, we obtain fidelities to the ideal state and quantum purities above 95\%, together with near-optimal quadrature squeezing. The ability to generate a wide variety of Wigner-negative states with mature semiconductor technology opens promising opportunities for continuous-variable optical quantum technologies.
\end{abstract}
	
\flushbottom
\maketitle

\thispagestyle{empty}

Encoding quantum information on light is a promising approach for scaling quantum information processing, benefiting from low decoherence and natural interconnectivity over long distances. Quantum information can be encoded on single photons using their degrees of freedom such as polarization or path (discrete variables (DV)), or on the quadratures of the electromagnetic field (continuous variable (CV)). Both approaches require optical nonlinearities at the single-photon level to implement universal quantum operations and achieve a quantum advantage in computation~\cite{hasnaoui_optical_2026, Lloyd1999, zhong_quantum_2020, Walschaers2021, Booth2022, knill_scheme_2001,Kok2007,Obrien2009, Lee2026}. Such nonlinearities can be realized probabilistically through measurement-induced interactions~\cite{fadrny_experimental_2024}, at the cost of substantial resource overhead~\cite{aghaee_rad_scaling_2025_xanadu, psiquantum2025manufacturable, knill_scheme_2001}, or deterministically by exploiting the intrinsic nonlinearity of single quantum emitters, such as trapped ions~\cite{blinov2004, barros2009, higginbottom2016}, atoms~\cite{ Bertet2002, mckeever2004, Daiss2019, Hacker2019, magro_deterministic_2023} or any low-noise solid-state artificial atom~\cite{Aharonovich2016}.  

Semiconductor quantum dots (QDs) embedded in optical microcavities have emerged as excellent sources of single~\cite{somaschi_near-optimal_2016, Tomm2021, ding_high-efficiency_2025, Loredo2026,huet2026_Bigpaper} and entangled photons \cite{wang2019XX, huet_deterministic_2025, schawrtz2016,cogan2023,coste_high-rate_2023,  Meng_DeterministicPhotonSource_2024} with clock rates of up to GHz~\cite{ding_high-efficiency_2025}. They have recently led to the development of the first optical quantum computing prototypes in the DV framework~\cite{maring_versatile_2024,pont_high-fidelity_2024,coste_high-rate_2023,huet_deterministic_2025,Uppu2021}. Very recently, the extremely low noise observed in QD-cavity devices has allowed for the measurement of the Wigner function of the single-photon state generated by the QD~\cite{huet2026_Bigpaper}. However, fundamentally, their strong single-photon nonlinearity~\cite{DeSantis2017} could also enable generating and carving out a large variety of non-Gaussian states of light~\cite{DeSantis2017, Ourjoumtsev2006,GeaBanacloche2013,Du2020,rosenblum_extraction_2016}.
For example, when controlled coherently, the coherence imprinted on the QD from the short driving laser pulse is mapped onto the optical field upon spontaneous emission, leading to the generation of vacuum--one-photon quantum states $|\Psi\rangle= \sqrt{1-p}\ket{0}+\sqrt{p}\exp(i\phi )\ket{1}$~\cite{loredo_generation_2019,Polacchi2024, DiMicco2025, MaillettedeBuyWenniger2024} with tunable single-photon population $p$ and a controlled phase $\phi$.

\begin{figure*}[]
    \centering
    \includegraphics[width=\linewidth]{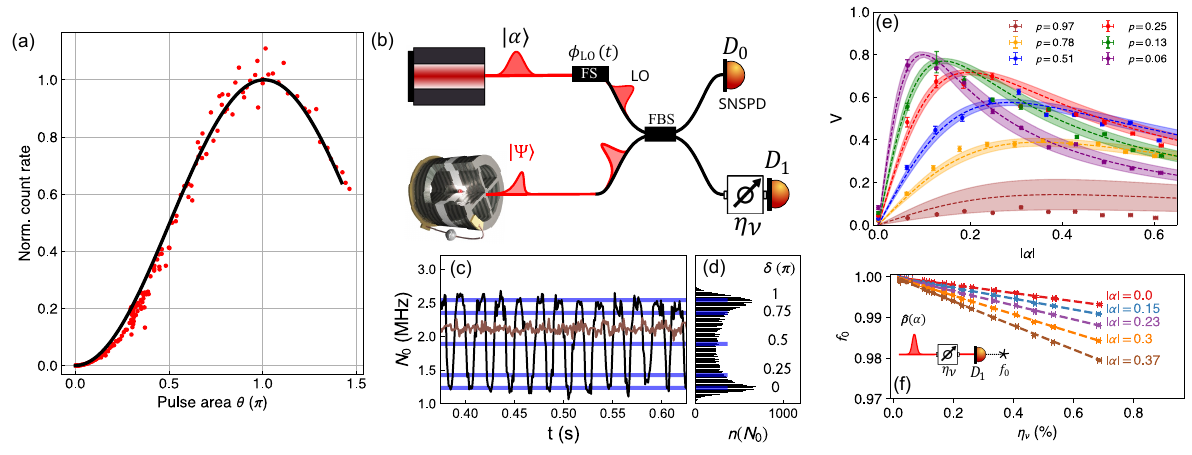}
    \caption{
    (a) Normalized single-photon count rate (data points) as a function of the pulse area $\theta$ evidencing the onset of Rabi oscillations with $p=\sin^2\theta/2$ (solid line).  
    (b) Schematic of the Wigner tomography setup: the measured field is displaced by LO on a fiber beamsplitter (FBS) with an SPNPD in each output ($D_0$ and $D_1$). The LO is derived from the excitation mode-locked laser and spectrally filtered with etalon-based spectral shaping; its phase $\phi_\mathrm{LO}$ is modulated with a fiber stretcher (FS). Phase-space reconstruction is performed using zero-photon counts on  $D_1$ with variable detection efficiency $\eta_\nu$ controlled by a calibrated optical attenuator. (c) Single-photon detected count rate $N_0$ on $D_0$ as a function of time, while scanning the LO phase (black $\theta=\pi/2$; brown $\theta=\pi$). (d)~Histogram of full count trace in panel (c) binned to 100 intensity bins, revealing the expected $\arccos (N_0/\mathrm{max}N_0)$ dependence, which is further binned into equidistant phase intervals (for clarity, we highlight in blue 5 phase bins; for Wigner functions we use 15 bins). (e)~Measured visibility of the field interference (symbols) agrees well with expected visibility for all measured vacuum-one-photon qubits as well as LO amplitudes (lines) with $M=0.68$. The shadowed areas represent the $1\sigma$-significant interval given by the uncertainty of~$p$. (f)~Example of zero-photon probability $f_0$ (symbols) characterizing the state emitted at $\pi$-excitation, measured for a range of detection efficiencies $\eta_\nu$ and multiple different phase-space coordinates $\alpha$ compared to theoretical expectations with the reconstructed state $\hat{\rho}$ (lines). 
    }
\label{fig:setup_QD}
\end{figure*}
	
Here, we perform Wigner tomography of these states using a phase-resolved weak-local-oscillator homodyne tomography scheme~\cite{zambra_experimental_2005}, resolving zero-photon detection events~\cite{huet2026_Bigpaper}. Correcting for optical losses up to the quantum dot to assess the QD emission property only, we observe a near-ideal Wigner function of the generated wavepackets. The Wigner functions evidence the continuous transition from the Gaussian vacuum state to the highly non-Gaussian single-photon Fock state, with observed high fidelity and quadrature squeezing close to theoretical estimates. 

We study a neutrally charged InGaAs QD embedded in a near-non-birefringent micropillar cavity cooled in a 5.5\,K closed-cycle cryostat. The QD transition is driven resonantly with 10\,ps-long laser pulses (repetition rate $f_\mathrm{rep}=80$\,MHz) aligned along the polarization axis of a cavity mode. The coherent control of the QD state creates a quantum superposition of the ground and excited states, which, upon spontaneous emission, leads to the generation of a quantum superposition of 0 and 1 photons $|\Psi\rangle= \sqrt{1-p}\ket{0}+\sqrt{p}\exp(i\phi )\ket{1}$~\cite{loredo_generation_2019}, emitted in a single temporal mode. This is evidenced by the  Rabi oscillations observed as a function of the pulse area $\theta$, reflecting the variation of the population $p=\sin^2(\theta/2)$ (Fig.~\ref{fig:setup_QD}a)~\cite{Giesz2016}. 
    
The emitted light field is separated from the laser by collecting the emission in a polarization perpendicular to the excitation~\cite{steindl2023cross}. For a $\theta=\pi$ excitation, the probabilities of collecting a photon from the cavity at the collection lens per excitation pulse are $B=37(2)\,\%$ and at the fiber output  $B_\mathrm{F}=13(1)\,\%$. The single-photon purity $1-g^{(2)}(0)=97.9(4)\,\%$ and indistinguishability $M_\mathrm{s}=95.9(2)\,\%$~\cite{Ollivier2021} at $\theta =\pi$  are determined using photon-correlation measurements, while the photon-number coherence $C_0=96.1(8)\,\%$ of the photonic wavepackets is measured with a self-homodyne detection technique varying $\theta$~\cite{loredo_generation_2019,MaillettedeBuyWenniger2024} (Supplementary).
	
We measure the Wigner function of the emitted light wavepackets, using a homodyne detection scheme (shown in Fig.~\ref{fig:setup_QD}b) in which the generated state described by $\hat{\rho}$ interferes with a weak local oscillator (LO)~\cite{laiho_probing_2010,nehra_state-independent_2019,huet2026_Bigpaper}. This interference on an intensity-balanced fiber-based beam splitter (FBS) effectively implements a tunable phase-space displacement operation $\hat{D}(\alpha)$, enabling the reconstruction of the Wigner function of $\hat{\rho}$ at a phase-space coordinate $\alpha$ given by $W(\alpha) = \tfrac{2}{\pi}\Tr\{\hat{\Pi}\hat{\rho}(\alpha)\}$, using the photon-number parity $\hat{\Pi}=(-1)^{\hat{n}}$ of the displaced quantum state $\hat{\rho}(\alpha)=\hat{D}(\alpha)\hat{\rho}\hat{D}^\dagger(\alpha)$, which can be measured from its photon-number statistics, enabling infering $\hat{\rho}$ if the displacement fidelity is known. The probed phase-space radial coordinate is given by the LO amplitude $|\alpha|$ and the azimuth by the relative phase between the two fields $\delta=\phi-\phi_\mathrm{LO}$. In the present experiment, we scan $\phi_\mathrm{LO}$ to deduce $\delta$ from oscillations in the single-count rate of the output detector $D_0$, with a detection efficiency $\eta_0$ (superconducting nanowire single-photon detector (SNSPD)). Neglecting the two-photon component of the QD emission, the detected clicks $N_0(t)\approx\frac{\eta_0 f_\mathrm{rep}}{2}\big[ |\alpha|^2 + B_\mathrm{F}p+ 2\sqrt{B_\mathrm{F} M(1-p)p|\alpha|^2} \cos(\phi - \phi_\mathrm{LO}) \big]$ oscillate with the scanned LO phase $\phi_\mathrm{LO}$ (example shown in Fig.~\ref{fig:setup_QD}c), with counts proportional to the intensities of individual fields (first two terms) and their interference (last term) scaling with the overlap of the signal-LO wavepackets $M$ and QD emission amplitude $\sqrt{p(1-p)}$. The normalized oscillations of ${N}_0$ are binned into 100 intensity occurrence intervals with a typical arccosine distribution (example in Fig.~\ref{fig:setup_QD}d). This distribution is binned into 15 equidistant phase intervals to extract the relative phase $\delta$ between 0 and $\pi$ bounded by the mirror symmetry of $|\Psi\rangle$. The time stamp of each of the detection events is then paired with time tags of the signal registered by an SNSPD detector $D_1$ at the second output of FBS to distribute the clicks of $D_1$ to the associated phase interval.

The pulse shaping of the LO pulses is tuned to maximize $M$ by monitoring the quantum interference between the single-photon state ($\theta=\pi$) and the LO, monitoring the cross-correlation signal between the two detectors $D_0$ and $D_1$ at the output of the FBS~\cite{lam_optimizing_2025}. We achieve an overlap of $M=68\pm1\,\%$, with measured temporal overlap at most $84\pm2\,\%$ limited by the complex wavepacket temporal profile of QD emission with imprinted quantum beats between its two excitonic states~\cite{Ollivier2020} (see Supplementary). During Wigner tomography, this displacement fidelity $M$ is then constantly monitored using the amplitude of $N_0(t)$ oscillations, resulting in a visibility $V=\frac{2\sqrt{ M(1-p)r_j}}{1+r_j}$, where $r_j=B_\mathrm{F}p/|\alpha|^2$ is the relative strength of the two fields. The measured $V$ in Fig.~\ref{fig:setup_QD}e is in good agreement with the expected value for $M=0.68(1)$, extracted from the homodyne correlation measured at $\theta=\pi$, showing that $|\Psi\rangle$ occupies an identical mode for all $p$. This behavior is expected as the emitted temporal wavepackets are emitted in the optical mode defined by the cavity. 
	
At each phase-space coordinate $\alpha$, we probe the photon-number statistics of $\hat{\rho}(\alpha)$ by measuring zero-photon probabilities $f_0(\eta_\nu, \alpha)$~\cite{zambra_experimental_2005,allevi_state_2009,Zhang2009} with $D_1$, registering the presence of at least one incident photon without photon-number resolution, after a calibrated attenuator effectively controlling the detection efficiency $\eta_\nu$. Fig.~\ref{fig:setup_QD}f shows $f_0$ under $\pi$-excitation for various LO strengths. The zero-photon detection probability links the detected click probability $f_\mathrm{det}(\eta_\nu,\alpha)$ with the photon-number probability distribution of $\hat{\rho}(\alpha)$
	\begin{align}
		f_0(\eta_\nu,\alpha) = \sum_n ^{N}(1-\eta_\nu)^n \langle n|\hat{\rho}(\alpha)|n\rangle =1 - f_\mathrm{det}(\eta_\nu,\alpha)\,.\nonumber
	\end{align}
The linearity of this relation enables the reconstruction of the full photon-number distribution of $\hat{\rho}(\alpha)$ up to $N$ photons from $N_\nu\ge N$ measurements performed at ideally equidistant $\nu$ detection efficiencies $\eta_\nu$~\cite{Kovtoniuk2026}. Repeating this procedure with various amplitudes and phases of LO enables phase-space reconstruction of the state $\hat{\rho}$, including its coherences, revealed by its field interferfence with LO. The state $\hat{\rho}$ is corrected during an iterative maximum likelihood (ML) reconstruction for optical losses up to the solid-state device~\cite{huet2026_Bigpaper, zambra_experimental_2005, lvovsky_iterative_2004} (see Supplementary).
\begin{figure}[!h]
    \centering
    \includegraphics[width=\linewidth]{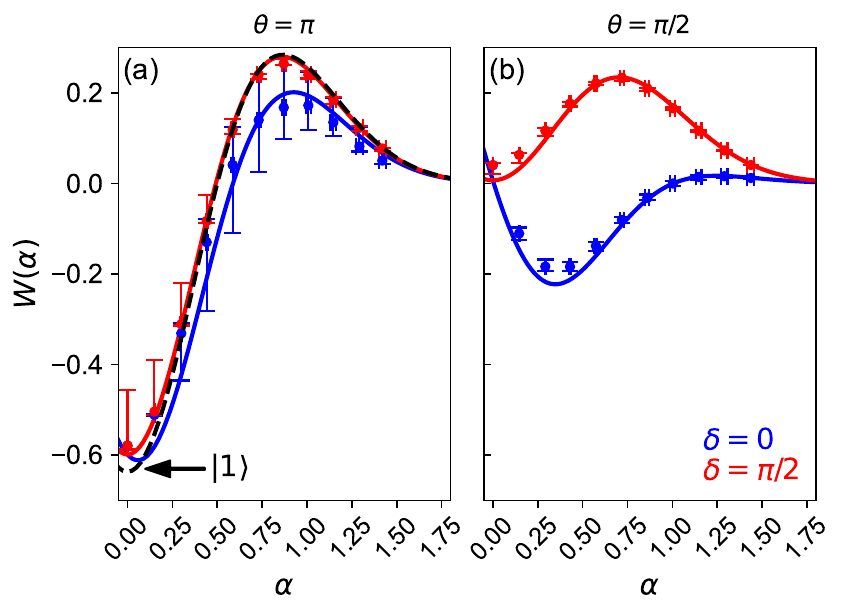}
    \caption{
    \textbf{Cross-sections of the reconstructed Wigner functions} of the Fock superposition state $\sqrt{1-p}|0\rangle+\sqrt{p}e^{i\phi}|1\rangle$ with (a) $p=0.97(3)$ (driven at $\theta=\pi$) and (b) $p=0.51(1)$ ($\theta=\pi/2$). The loss-corrected measured data (symbols) show excellent agreement with $W(\alpha)$ of the corresponding pure states (solid lines), including their Wigner negativity and phase-space asymmetry presented using cross-sections along two orthogonal phase-space directions (red $\delta=\pi/2$, blue $\delta=0$). The asymmetry in (b) corresponds to quadrature squeezing along $\delta=0$ and anti-squeezing along $\delta=\pi/2$. In (a), the state is compared to the perfect $|1\rangle$ (black).
    }
    \label{fig:Wigner_crosssection}
\end{figure}
	
	{
      Fig.~\ref{fig:Wigner_crosssection}a shows the measured loss-corrected Wigner function of the photonic state triggered by the $\pi$-pulse excitation. The corresponding photon-number distribution for zero displacement shows a dominant single-photon population of $95.1\,\%^{+3.2\,\%}_{-1.7\,\%}$, around $3\,\%$ of vacuum, and $1\,\%$ of two-photon population~\cite{MaillettedeBuyWenniger2024} that unavoidably originates from the two-level system dynamics under finite pulse excitation~\cite{Hanschke2018, Ollivier2021}. The Wigner function evidences a negativity, a resource for CV quantum advantage~\cite{Walschaers2021}, of $\mathrm{min}(W(\alpha))=-0.579^{+0.122}_{-0.009}$ and Wigner-negative volume~\cite{Kenfack2004} $\mathcal{N}_W=\int_{W<0} |W(\alpha)|\mathrm{d^2\,\alpha}=0.385^{+0.001}_{-0.043}$.
	These values closely approach the ideal single-photon limits  $\mathrm{min}(W(\alpha))=-0.637$ and $\mathcal{N}_W=0.426$, and are only marginally reduced by residual vacuum and two-photon contributions. 
    The coherent reexcitation dynamics, leading to the two-photon component, induces reminiscent photon-number coherence and result in a rotational phase space asymmetry, as shown by the reconstruction along two orthogonal directions in the phase space in Fig.~\ref{fig:Wigner_crosssection}(a).    Fig.~\ref{fig:Wigner_crosssection}(b) shows similar data for two orthogonal phase-space directions for $\theta=\pi/2$, evidencing this time strong phase dependence as expected for a balanced quantum superposition of zero and one photon.

\begin{figure}[]
	\centering
    \includegraphics[width=1\linewidth]{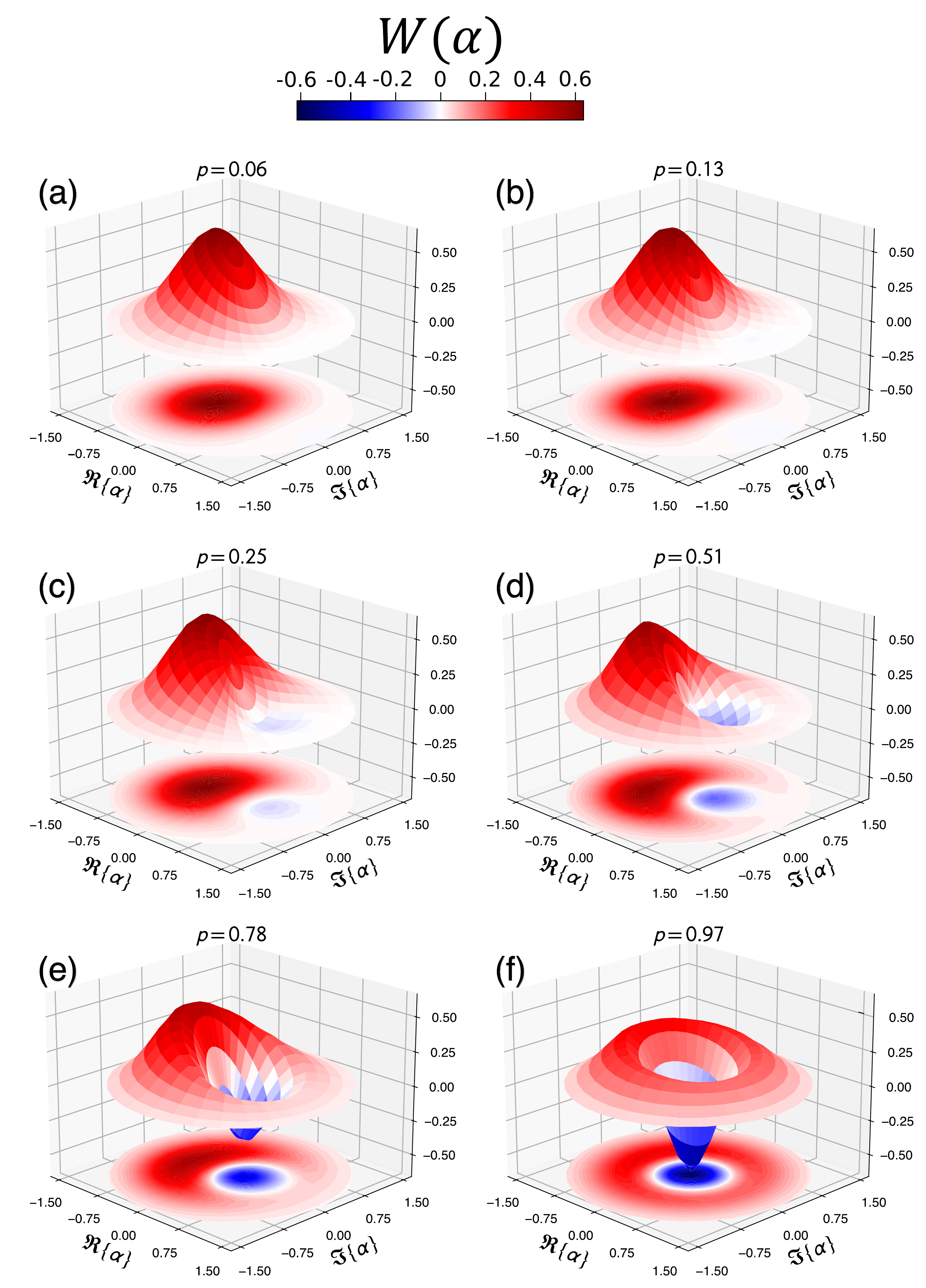}
	\caption{\textbf{Propagation-loss-corrected Wigner functions for vacuum-one-photon superposition states $\sqrt{1-p}|0\rangle+\sqrt{p}|1\rangle$}. (a)-(f) The point-by-point reconstructed Wigner functions $W(\alpha)$ for increasing single-photon population $p$ reveal the emergence of Wigner-function negativity. Additionally, we observe a $p$-dependent phase-space asymmetry (better visible in the 2D plot under the 3D image), quantified in Fig.~\ref{fig:WignerFunction_extractedparam}. 
	}    \label{fig:WignerFunction2D}
\end{figure}

\begin{figure}[]
	\centering
    \includegraphics[width=1\linewidth]{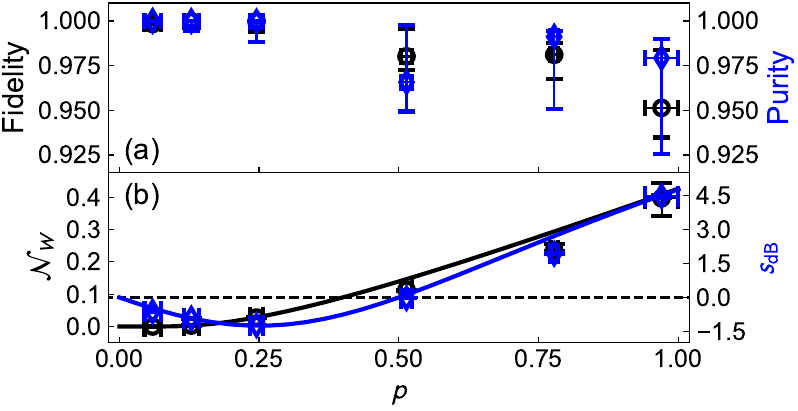}
	\caption{\textbf{Non-Gaussianity and state purity of the generated states}. We quantify the Wigner negativity, which is equivalent to non-Gaussianity if the state is pure~\cite{Hudson1974}. (a) The reconstructed quantum state purity (blue) and its fidelity to the target vacuum-one-photon qubit (black). (b) Data points extracted from Fig.~\ref{fig:WignerFunction2D} quantify the photonic-qubit Wigner-negativity volume $\mathcal{N}_W$ (black) and the quadrature squeezing parameter $s_\mathrm{dB}$ (blue). Solid lines indicate the expected behavior for an ideal vacuum-one-photon qubit.}  \label{fig:WignerFunction_extractedparam}
\end{figure}
    
Fig.~\ref{fig:WignerFunction2D} presents the tomographic reconstruction of six light wavepackets corresponding to different pulse areas $\theta$, hence different $p$. The tomography of all states together was performed in a single, realignment-free experimental run, where we probe each state sequentially with LO of 11~different strengths. During this time, the device emits a pure superposition state into a single well-defined mode with a stable phase relation between the vacuum and a single photon, enabling the reconstruction of the Wigner function. In all cases, the Wigner reconstruction provides excellent agreement with the ideal target states, revealing both the Wigner-negative and phase-sensitive regions. Fig.~\ref{fig:WignerFunction_extractedparam}a  reports the purities and fidelities~\cite{Jozsa1994} of the reconstructed states, evidencing values above $95\,\%$. These results evidence the extremely low noise in our QD devices, enabled by high-purity semiconductor growth, stable electrical control, and pure dephasing reduction provided by Purcell enhancement~\cite{huet2026_Bigpaper}.

Fig.~\ref{fig:WignerFunction_extractedparam}b reports the quantities directly linked to state nonclassicality extracted from the loss-corrected Wigner tomography. At low $p$, the emission quadrature variance $\langle \Delta^2 x_\delta\rangle$ exhibits squeezing along the $\delta$ direction, reaching at $p=0.25$ the maximal quadrature squeezing~\cite{Lvovsky2015} of $s_\mathrm{dB}=10\log_{10} (4\langle \Delta^2 x_\delta\rangle) = 1.244^{+0.043}_{-0.003}\,\mathrm{dB}$ or $24.8^{+0.2}_{-1.0}\,\%$ below the shot noise level — a value very close to the optimal of $25\,\%$ for an ideal vacuum-single-photon qubit~\cite{Walls1981, Carmichael1985}, reduced mostly by a small %incoherent mixture with effective 
two-photon component, originating from the QD reexcitation dynamics~\cite{Fischer2017} (see Supplementary).
Such a high degree of squeezing contrasts with the value measured in the pioneering work of Schulte et al., with a value about 40\,\% of the ideal one, evidencing stronger decoherence for QDs in bulk~\cite{schulte_quadrature_2015}.

Fig.~\ref{fig:WignerFunction_extractedparam}(b) also shows the reconstructed Wigner negativity~\cite {Genoni2013} which increases with $p$: the state is effectively tuned between a %fully Gaussian 
vacuum state (at $p=0$ with $\mathcal{N}_W=0$) and a single-photon state with extracted $\mathcal{N}_W=0.385^{+0.001}_{-0.043}$.

%=================================================
%\section*{Conclusions}
%=================================================

\vspace{0.3 cm}
To conclude, we have experimentally reconstructed the Wigner function of light wavepackets emitted by a semiconductor QD under resonant excitation with continuous control over the non-Gaussianity and negativity of the emitted states. The measured states show near-unity purity and fidelity to the ideal states owing to the very low noise of our system~\cite{huet2026_Bigpaper}.

\noindent  This demonstration relies on a loss correction up to the emitter, since high photon-number coherence QD devices require operating with a neutral QD that evidences two linear optical dipoles, which limits the source efficiency to 50\,\% in an unpolarized cavity~\cite{Ollivier2020}. Such device efficiency could be significantly increased to obtain Wigner negativity at the device output by polarization-selective, high-quality factor cavities, reducing laser-rejection induced losses \cite{Wang2019}.

\noindent These results open the path to the generation of more complex photonic states such as cat states, following demonstrations with trapped atoms~\cite{Hacker2019} using here semiconductor integrated QD-cavity devices operating at up to GHz rates.

\bibliography{Wigner_paper_biblio}

\section*{Acknowledgements}

The authors acknowledge William Hease, Anton Pishchagin, Sebastien  Boissier, and Thi Huong Au for providing the sample for this study, and Loïc Lanco, Samyak Pratyush Prasad, Shane Mansfield, and Jean Senellart for fruitful discussions. HL thanks Alexei Ourjoumtsev for his stimulating discussions, which led to significant improvements in the method. This work was conducted within the research program of the QDLight joint laboratory (C2N/Quandela). It was partially supported by the Paris Ile-de-France Région in the framework of DIM QUANTIP, the French Defense ministry - Agence de l'innovatio de défense, the European Union’s Horizon CL4 program under the grant agreement 101135288 for EPIQUE project, the Plan France 2030 through the projects ANR22-PETQ-0011, ANR-22-PETQ-0006, and ANR-22-PETQ-0013, and a public grant overseen by the French National Research Agency as part of the ”Investissements d’Avenir” programme (Labex NanoSaclay, reference: ANR10LABX0035). This work was carried out within the C2N micro-nanotechnologies platforms and was partially supported by the RENATECH network and the General Council of Essonne.

\end{document}